\documentclass[prd,twocolumn,showpacs,preprintnumbers,amsmath,amssymb,floatfix]{revtex4}

\usepackage{latexsym}
\usepackage{graphicx}
\usepackage{subfigure}

\newcommand{\mev}{\textrm{ MeV}}

\newcommand{\be}{\begin{equation}}
\newcommand{\ee}{\end{equation}}
\newcommand{\ba}{\begin{eqnarray}}
\newcommand{\ea}{\end{eqnarray}}

\newcommand{\nn}{\nonumber}

\begin{document}

\title{On the role of meson loops in the
$f_0(1710)$ decay into $V\gamma$}

\author{
 H.~Nagahiro$^1$, L.~Roca$^2$, E.~Oset$^3$ and B.~S.~Zou$^4$}

\affiliation{
$^1$Research Center for Nuclear Physics (RCNP), Ibaraki, Osaka
  567-0047, Japan \\
 $^2$Departamento de F\'{\i}sica. Universidad de Murcia. E-30071
Murcia.  Spain \\
$^3$Departamento de F\'{\i}sica Te\'orica and IFIC,
Centro Mixto Universidad de Valencia-CSIC,\\
Institutos de
 Investigaci\'on de Paterna, Aptdo. 22085, 46071 Valencia, Spain\\
$^4$Institute of High Energy Physics and Theoretical Physics Center
for Science Facilities, CAS, Beijing 100049, China }

\date{\today}

\begin{abstract}

 We investigate the relevance of the meson loops  in the $f_0(1710)$
scalar meson decay into one photon and one vector meson, ($\rho$, $\omega$
and $\phi$). In particular we estimate the size of the loops coming from
the decay of the $f_0(1710)$ into two pseudoscalar mesons, containing
three pseudoscalar mesons in the loop or two pseudoscalar and one vector
meson. The results, despite having large uncertainties, manifest that the
contribution of the meson loops to these radiative decays is  quite
relevant  and should be taken into account by the theoretical
calculations which use this observables as a test of the possible glueball
nature of this resonance.

\end{abstract}

\maketitle

\section{Introduction}

One of the striking features of QCD is the possibility of getting
meson states formed from bound states of gluons, glueballs, and some
lattice calculations hint at their existence \cite{glue} with masses
around $1.3\sim 1.7$ GeV. In Refs. \cite{weingarten,chanowitz,chao}
the $f_0(1710)$ scalar meson is suggested as a possible glueball. In
\cite{chanowitz} it is argued that chiral symmetry has as a
consequence that the glueball decays predominantly to $s \bar{s}$
and the ratio of decay amplitudes  $G ~\to~\bar{s}s$ to $G
~\to~\bar{u}d+\bar{d}d$ is of the order of $m_s/2\hat{m}\simeq
m_K^2/m_\pi^2$, with $m_K$ and $m_\pi$ the kaon and pion masses;
while in \cite{chao} $G ~\to~\bar{q}q\bar qq$ is claimed to be the
dominant mechanism to reproduce the BES observed $\pi\pi/\bar KK$
ratio $\sim 0.41$ for $f_0(1710)$ \cite{Ablikim:2006db}. In a recent
work, using multiple meson-meson channels to build up the scalar
resonances \cite{Albaladejo:2008qa}, it is found that the
$f_0(1710)$ scalar meson couples very strongly to $\bar{s}s$ and
then it is suggested that it could be indeed a glueball. Yet, one of
the tests which is supposed to be very relevant to find out the
nature of the meson resonances is their radiative decay
\cite{Close:2002sf,DeWitt:2003rs}. Indeed, in \cite{Close:2002sf} a
thorough investigation of the radiative decay of the $f_0(1710)$
into $\rho \gamma$ and $\phi \gamma$ is done and large decay widths
are obtained depending on the amount of mixing with other non
glueball components, as suggested in \cite{Am95,CK02a}.

    The importance of the radiative decay to investigate the nature of
resonances has also been pointed out in
\cite{Kalashnikova:2005zz,Ivashyn:2007yy,Nagahiro:2008mn} in
connection with the scalar $f_0(980)$ and $a_0(980)$ resonances, which in
chiral unitary theories appear as dynamically generated
\cite{Oller:1997ti,Kaiser:1998fi,Markushin:2000fa,Nieves:1999bx,Oller:1998hw,Oller:1998zr}.
With the assumption that the   $f_0(980)$ and $a_0(980)$ resonances are
dynamically generated, hence basically molecules of the $\pi \pi$ and $K
\bar{K}$ coupled channels, the radiative decay of these resonances is done by
photon emission from the meson components, which technically appear as loop
contributions. In
 \cite{Kalashnikova:2005zz} the important contribution of the $K \bar{K}$ loops
 to the radiative decay of those scalar resonances into $\rho^0 \gamma$ and
 $\omega \gamma$ was evaluated. In \cite{Ivashyn:2007yy} additional pion loops
 were considered
 and in \cite{Nagahiro:2008mn} the contribution of intermediate vector
 meson channels was also taken into account, which in the case of the $a_0(980)$
 radiative decay turned out to be important.

    Given the fact that the loops gave sizable contributions to the radiative
 decay widths of the $f_0(980)$ and $a_0(980)$ resonances, it looks important
 that a determination of their strength for higher mass scalar mesons is carried
 out.  This is the purpose of the present paper where we determine for the case
 of the $f_0(1710)$ the
 contribution of the loops which were evaluated in \cite{Nagahiro:2008mn} for
 the case of the $f_0(980)$ and $a_0(980)$ resonances.
 Unlike the case of the $f_0(980)$ and $a_0(980)$ resonances, where the couplings
 of the resonances to the different channels is provided by the chiral unitary
 theories, in the present case we have to resort to phenomenology. The rest of
 the information needed for the evaluation follows closely the steps of
\cite{Nagahiro:2008mn}.  The choice of the $f_0(1710)$ resonance to
make this study is also motivated because its experimental study is
feasible at BESIII in near future and plans to do it are under way
with partial wave analysis tools ready \cite{Dulat:2005in}.

\section{Formalism}

According to the Particle Data Group (PDG) \cite{pdg2007online}, the
$f_0(1710)$ scalar resonance decays mostly into two pseudoscalar
mesons, and among the pseudoscalar mesons the $K\bar K$ channel is
the dominant one. From the experimental value $\Gamma_{K\bar
K}/\Gamma_{\textrm{total}}=0.38^{+0.09}_{-0.19}$
\cite{pdg2007online,Longacre:1986fh} and from the expression of the
decay of the $f_0(1710)$ into $K\bar K$, \be \Gamma_{K\bar
K}=\frac{\beta}{16\pi M_f}|g_{K\bar K}|^2, \ee
where $M_f$ is the mass of the  $f_0(1710)$ and
$\beta=\sqrt{1-4m_K^2/s}$,
the coupling of the $f_0(1710)$ to the $K\bar K$ isospin $I=0$
state results $g_{K\bar K}\simeq 2350\pm 500\mev$.
It is worth noting that this value compares very well with
that obtained with the
coupled channel analysis of
\cite{Albaladejo:2008qa,Albaladejoprivate},
$g_{K\bar K}\simeq 2862\pm 440\mev$.

\begin{figure*}[hbt]
  \begin{center}
\resizebox{0.7\linewidth}{!}{%
  \includegraphics{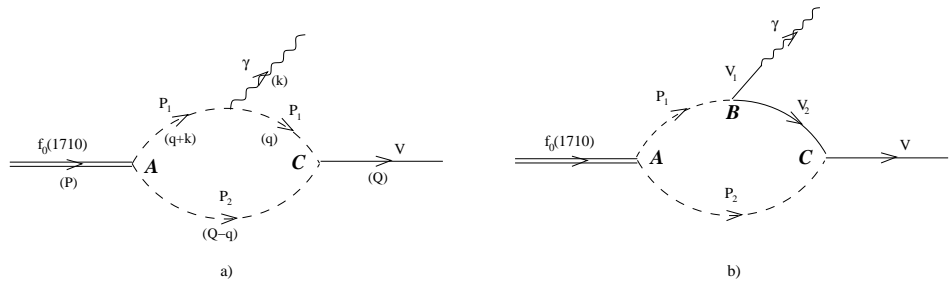}
}
\caption{Feynman diagrams contributing to the $f_0(1710)$
 radiative
decay. The variables within brackets represent the momenta; $P_1$,
$P_2$, the different allowed kaons and $V$, $V_1$, $V_2$, the
different vector mesons. $A$, $B$ and $C$ are coefficients explained
in the text.}
 \label{fig:diag}
 \end{center}
\end{figure*}

The idea to follow in order  to evaluate the radiative decay width is
similar to the procedure used in ref.~\cite{Nagahiro:2008nz} which we
summarize here adapted to the present case. Once we have the value of the
coupling of the $f_0(1710)$ to the $K\bar K$, the philosophy is  to consider the
transition from the  $f_0(1710)$ mesons to the   $K\bar K$ states at one
loop and then attach the photon to the possible allowed lines,
considering that a vector meson  in the final state has to be produced.
By using arguments of gauge invariance, it can be shown that we only
need to evaluate the diagrams depicted in fig.~\ref{fig:diag}, despite the fact
that other
diagrams are in principle needed like, {\it e.g.}, those with a photon
emitted from the initial and final vertices.

The channels we will consider are $f_0(1710)\to\rho^0\gamma$,
$f_0(1710)\to\omega\gamma$ and $f_0(1710)\to\phi\gamma$.
In fig.~\ref{fig:diag}
$P_1$, $P_2$, represent the kaons in the loop
 and $V_1$, $V_2$, vector
mesons.

\begin{table*}[hbt]
\begin{center}
\begin{tabular}{c|c|c|c}
decay & $P_1P_2$ & $A$ & $C$ \\\hline\hline
$f_0(1710)\to\rho^0\gamma$ & $K^+K^-$     & $-1/\sqrt{2}$ & $-1/\sqrt{2}$  \\
                   & $K^-K^+$     & $-1/\sqrt{2}$ & $ 1/\sqrt{2}$  \\ \hline
$f_0(1710)\to\omega\gamma$ & $K^+K^-$     & $-1/\sqrt{2}$ & $-1/\sqrt{2}$  \\
                   & $K^-K^+$     & $-1/\sqrt{2}$ & $ 1/\sqrt{2}$  \\ \hline
$f_0(1710)\to\phi\gamma$   & $K^+K^-$     & $-1/\sqrt{2}$ & $ 1$           \\
                   & $K^-K^+$     & $-1/\sqrt{2}$ & $-1$           \\
\end{tabular}
\end{center}
\caption{Coefficients $A$ and $C$ for type-a diagrams}
\label{tab:coeff_type_a}
\end{table*}

\begin{table*}[htb]
\begin{center}
\begin{tabular}{c|r|c|c|c}
decay & $P_1P_2V_2V_1$ & $A$ & $B$ & $C$ \\\hline\hline
$f_0\to\rho^0\gamma$ & $K^+K^-K^{*+}\rho^0$     & $-1/\sqrt{2}$ & $1/\sqrt{2}$ & $1/\sqrt{2}$ \\
                     & $\omega$                 & $$            & $1/\sqrt{2}$ & $$ \\
                     & $\phi$                   & $$            & $1$          & $$ \\  \cline{2-5}
                     & $K^-K^+K^{*-}\rho^0$     & $-1/\sqrt{2}$ & $1/\sqrt{2}$ & $1/\sqrt{2}$ \\
                     & $\omega$                 & $$            & $1/\sqrt{2}$ & $$ \\
                     & $\phi$                   & $$            & $1$          & $$ \\  \cline{2-5}
                     & $K^0\bar{K^0}K^{*0}\rho^0$& $-1/\sqrt{2}$ & $-1/\sqrt{2}$ & $-1/\sqrt{2}$ \\
                     & $\omega$                 & $$            & $1/\sqrt{2}$ & $$ \\
                     & $\phi$                   & $$            & $1$          & $$ \\  \cline{2-5}
                     & $\bar{K^0}K^0\bar{K}^{*0}\rho^0$& $-1/\sqrt{2}$ & $-1/\sqrt{2}$ & $-1/\sqrt{2}$ \\
                     & $\omega$                 & $$            & $1/\sqrt{2}$ & $$ \\
                     & $\phi$                   & $$            & $1$          & $$ \\  \hline
$f_0\to\omega\gamma$ & $K^+K^-K^{*+}\rho^0$     & $-1/\sqrt{2}$ & $1/\sqrt{2}$ & $1/\sqrt{2}$ \\
                     & $\omega$                 & $$            & $1/\sqrt{2}$ & $$ \\
                     & $\phi$                   & $$            & $1$          & $$ \\  \cline{2-5}
                     & $K^-K^+K^{*-}\rho^0$     & $-1/\sqrt{2}$ & $1/\sqrt{2}$ & $1/\sqrt{2}$ \\
                     & $\omega$                 & $$            & $1/\sqrt{2}$ & $$ \\
                     & $\phi$                   & $$            & $1$          & $$ \\  \cline{2-5}
                     & $K^0\bar{K^0}K^{*0}\rho^0$& $-1/\sqrt{2}$ & $-1/\sqrt{2}$ & $1/\sqrt{2}$ \\
                     & $\omega$                 & $$            & $1/\sqrt{2}$ & $$ \\
                     & $\phi$                   & $$            & $1$          & $$ \\  \cline{2-5}
                     & $\bar{K^0}K^0\bar{K}^{*0}\rho^0$& $-1/\sqrt{2}$ & $-1/\sqrt{2}$ & $1/\sqrt{2}$ \\
                     & $\omega$                 & $$            & $1/\sqrt{2}$ & $$ \\
                     & $\phi$                   & $$            & $1$          & $$ \\  \hline
$f_0\to\phi\gamma$   & $K^+K^-K^{*+}\rho^0$     & $-1/\sqrt{2}$ & $1/\sqrt{2}$ & $1$ \\
                     & $\omega$                 & $$            & $1/\sqrt{2}$ & $$ \\
                     & $\phi$                   & $$            & $1$          & $$ \\  \cline{2-5}
                     & $K^-K^+K^{*-}\rho^0$     & $-1/\sqrt{2}$ & $1/\sqrt{2}$ & $1$ \\
                     & $\omega$                 & $$            & $1/\sqrt{2}$ & $$ \\
                     & $\phi$                   & $$            & $1$          & $$ \\  \cline{2-5}
                     & $K^0\bar{K^0}K^{*0}\rho^0$& $-1/\sqrt{2}$ & $-1/\sqrt{2}$ & $1$ \\
                     & $\omega$                 & $$            & $1/\sqrt{2}$ & $$ \\
                     & $\phi$                   & $$            & $1$          & $$ \\  \cline{2-5}
                     & $\bar{K^0}K^0\bar{K}^{*0}\rho^0$& $-1/\sqrt{2}$ & $-1/\sqrt{2}$ & $1$ \\
                     & $\omega$                 & $$            & $1/\sqrt{2}$ & $$ \\
                     & $\phi$                   & $$            & $1$          & $$ \\  \hline

\end{tabular}
\end{center}
\caption{Coefficients $A$, $B$ and $C$
for the different allowed type-b diagrams.}
\label{tab:coeff_type_b}
\end{table*}

In
tables~\ref{tab:coeff_type_a} and \ref{tab:coeff_type_b}
 we show the different
allowed
$P_1P_2V_1V_2$ particles of the diagrams in fig.~\ref{fig:diag},
together with the corresponding coefficients for each channel to be
explained latter on. In fig.~\ref{fig:diag}, $P$, $q$, $k$ and $Q$
represent the momentum of the different lines that will be used in
 the evaluation of the loop function.

Next we show how gauge invariance is invoked in order
to simplify the calculations.
We follow a
similar procedure as done in
refs.~\cite{Close:1992ay,Marco:1999df} in the evaluation of the
radiative $\phi$ decay, in refs.~\cite{Roca:2006am,Nagahiro:2008zz}
 for the radiative axial-vector meson decays and in
 ref.~\cite{Nagahiro:2008mn} for the radiative decays of the
 $f_0(980)$ and $a_0(980)$ scalar mesons.

We can write the general expression of the amplitude for
the radiative decay of the $f_0(1710)$
meson into  a vector meson and a photon ($f_0(1710)\to V\gamma$)
as
  \be
 T={\epsilon_V}_\mu \epsilon_\nu T^{\mu\nu}
 \label{eq:Tgeneral}
 \ee
with
\be
T^{\mu\nu}=a\, g^{\mu\nu} + b\, Q^\mu Q^\nu + c\, Q^\mu k^\nu
 +d\, k^\mu Q^\nu + e\, k^\mu k^\nu
 \label{eq:Tmunuterms}
\ee
where
 $Q$ is the final vector meson momentum  and $k$
the photon momentum.
In Eq.~(\ref{eq:Tgeneral}), $\epsilon_V$ and $\epsilon$ are the
vector  meson and photon polarization vectors respectively.
Due to the Lorenz condition, ${\epsilon_V}_\mu
Q^\mu=0$, ${\epsilon}_\nu k^\nu=0$, the $a$ and $d$ terms
are the only ones  in
Eq.~(\ref{eq:Tmunuterms}) which do not vanish.
On the other hand,
 gauge invariance implies that $T^{\mu\nu}k_\nu=0$,
from where one gets
\be a=-d\,Q\cdot k.
\label{eq:ad}
\ee
Therefore the amplitude gets the general form
 \be
 T=-d(Q.k \,g^{\mu\nu}-k^\mu Q^\nu){\epsilon_V}_\mu
  \epsilon_\nu .
 \label{eq:Tgeneral2}
 \ee
This implies that we only need to evaluate those diagrams
contributing to the $d$-term, which are
those having a final structure $k^\mu
Q^\nu$.
  The advantage to evaluate only
the $d$ coefficient is that
only the loop diagrams of fig.~\ref{fig:diag} contribute
since other diagrams, like those involving photon couplings to the
vertices which are necessary to fulfill gauge invariance, do not give
contribution to the $d$ coefficient
 \cite{Close:1992ay,Marco:1999df,Roca:2006am,Nagahiro:2008mn}.

In  ref.~\cite{Nagahiro:2008mn} it was shown that the $d$ coefficients are
finite for the type-a diagrams of fig.~\ref{fig:diag}. By doing the same
calculation as in ref.~\cite{Nagahiro:2008mn} (see that reference for the
explicit expressions of the Lagrangians needed) the $d-$coefficient for
the type-a diagrams is given by the finite expression:

\be
d_a=AC\,Q_1\,g_{K\bar K} \sqrt{2}e\frac{M_V G_V}{f^2}\frac{1}{8\pi^2}
\int_0^1dx\int_0^xdy\frac{(1-x)y}{s+i\varepsilon}
\label{eq:da}
\ee
\noindent
where $s=Q^2x(1-x)+2Q\cdot k(1-x)y+(m_2^2-m_1^2)x-m_2^2$ and
$A$ are coefficients given in table~\ref{tab:coeff_type_a}
needed to relate the $f_0(1710)P_1P_2$ coupling in charge basis with those
in isospin basis, $g_{K\bar K}$.
In Eq.~(\ref{eq:da}), $f$ is the pion decay constant for which we
take the same value as in ref.~\cite{Nagahiro:2008mn},
$Q_1$ is the sign of the
charge of the
$P_1$ pseudoscalar meson, $e$ is taken
positive, $G_V$ is the $VPP$ coupling constant
defined in \cite{Ecker:1988te} and for which we use the numerical
value $G_V=55\pm 5\mev$ from ref.~\cite{Palomar:2003rb},
$m_1$($m_2$) is the mass of the
$P_1$($P_2)$ pseudoscalar meson, $M_V$ the mass of the final vector meson
and $C$ are coefficients coming from the
vector-pseudoscalar-pseudoscalar ($VPP$) Lagrangian of
ref.~\cite{Nagahiro:2008mn} after performing the trace of
the matrix and
which depend on the particular $P_1$, $P_2$ and $V$ particles
(specifically, $C$ is the coefficient of $\langle
V^\mu[\partial_\mu P,P]\rangle$ of the $VPP$ Lagrangian
 defined as $C(P_1\partial P_2-P_2\partial P_1)$). The different
 $A$, $C$, coefficients are given in
table~\ref{tab:coeff_type_a}.

In ref.~\cite{Nagahiro:2008mn} it was shown that the type-b loops of
fig.~\ref{fig:diag} played a relevant role in the radiative decays into
$V\gamma$ of the $f_0(980)$ and $a_0(980)$ scalar resonances. Thus we also
evaluate the contribution of this kind of loop. The Lagrangians used for
the vertices containing fields other than the $f_0(1710)$ are the same as
in ref.~\cite{Nagahiro:2008mn}. In that reference it was also shown that,
despite the type-b loop being  apparently quadratically divergent, it
could be reduced to a logarithmic divergence  which can be easily
identified and regularized by expressing it in terms of the two
pseudoscalar mesons loop function. Actually the expression of the
$d$-coefficient for the type-b mechanisms can be split into a convergent
part,
\ba d_b^\textrm{con}&=& -ABC\,g_{K\bar K} \frac{N_BN_C G'^2 F_V}{2
M_1}e\lambda_{V_1} \frac{1}{32\pi^2}\nn\\&\times&
\int_0^1dx\int_0^xdy\frac{1}{s'+i\varepsilon}(Q^2(1-x)^2-M_2^2)
\label{eq:dbconv}
\ea
 and a divergent one,
   \be
d_b^\textrm{div}= -ABC\,g_{K\bar K}\frac{N_BN_C G'^2 F_V}{4
M_1}e\lambda_{V_1} G(P^2,m_1,m_2),
\label{eq:dbdiv}
\ee
written in terms of the two pseudoscalar mesons loop function,
$G(P^2,m_1,m_2)$, which can be properly regularized either with a cutoff
\cite{Oller:1997ti} or  with dimensional regularization
\cite{Oller:1998zr,Oller:1998hw}. Loops with vector mesons are regularized in
\cite{Bruns:2004tj} with dimensional regularization and in
\cite{Nagahiro:2008mn} also with a cutoff. The results obtained with the 
cutoff method are very similar with those using the procedure outlined before,
which is the one we follow here.

In Eq.~(\ref{eq:dbconv}) and
(\ref{eq:dbdiv}), $s'=Q^2x(1-x)+2Q\cdot k(1-x)y+(m_2^2-M_2^2)x+
(M_2^2-m_1^2)y-m_2^2$; $M_1$($M_2$) is the mass of the $V_1$($V_2)$ vector
meson; $\lambda_{V}$  is $1$, $1/3$, $-\sqrt{2}/3$ for $V=\rho$, $\omega$,
$\phi$ respectively and $F_V$ is a coupling constant in the
normalization of \cite{Ecker:1988te} for which we use the value
$F_V=156\pm 5\mev$ \cite{Palomar:2003rb}. The coefficient $A$ has the same
meaning as in the type-a loop case and $B$ is the coefficient of the
$P_1V_1V_2$ vertex obtained after performing the $SU(3)$ trace in
$\langle V V P \rangle$ (see ref.~\cite{Nagahiro:2008mn}) defined as
$B\,P_1\bar{V_1}\bar{V_2}$. Analogously, $C$ is the coefficient coming
from the $P_2 V_2 V$ vertex defined as  $C\,P_2\bar{V} V_2$ from the
resulting expression after taking the trace in $\langle V V P \rangle$.
The $N_B$ and $N_C$ coefficients are normalization factors for the $B$ and
$C$ $VVP$ vertices (with coupling $G'$) such that the $V\to P\gamma$
decays agree with experiment, as explained in ref.~\cite{Nagahiro:2008mn}.

The total amplitude for the radiative decay process is
given by Eq.~(\ref{eq:Tgeneral2}) where
$d=d_a+d_b^\textrm{con}+d_b^\textrm{div}$ from
Eqs.~(\ref{eq:da}), (\ref{eq:dbconv}) and
(\ref{eq:dbdiv}).

The radiative decay width of the $f_0(1710)$ resonance
 into a vector meson and a photon is given,
in the narrow resonance limit, by
\ba \nn
&&\Gamma(M_f,M_V) = \frac{|\vec{k}|}{8\pi M_f^2} \Sigma|T|^2\,
 \theta(M_f-M_V)\\
&&=
\frac{M_f^3}{32\pi}\left(1-\frac{M_V^2}{M_f^2}\right)^3 |d|^2\,
 \theta(M_f-M_V),
\ea
where $\theta$ is the step function.

The finite widths of the $f_0(1710)$
resonance and the final vector meson are taken into account
by folding the previous expression
with their corresponding mass distributions, in a similar way as
explained in refs.\cite{Nagahiro:2008mn,Nagahiro:2008nz}.

\begin{table*}
\begin{center}
\begin{tabular}{|c|c|c|c|}
\hline
& loop a & loop b & total \\ \hline\hline

$f_0(1710)\to\rho^0\gamma$ &  4.3 & 75.5  & $100\pm 40$  \\ \hline

$f_0(1710)\to\omega\gamma$ & 4.5 &  2.34  & $3.3\pm 1.2$ \\ \hline

$f_0(1710)\to\phi\gamma$   & 6.9 &  2.9   & $15\pm 5$ \\ \hline

\end{tabular}
\end{center}
\caption{
Contribution of the different meson loops to the radiative
decay widths. All the units are in KeV.}
\label{tab:results1}
\end{table*}

\section{Results}

In  table~\ref{tab:results1} we show our results for the contributions
 of the
type-a and -b loops to the radiative decay widths under
consideration.
The theoretical
errors quoted in our final results  have been obtained by doing a
Monte-Carlo sampling of the parameters of the model within their
uncertainties. It is worth stressing that our results must be viewed only
as qualitative since we are not considering
loops coming from decay channels of the $f_0(1710)$ other than $K\bar K$.

From the results in table~\ref{tab:results1} we can see
that the contribution of the loops considered in the present work to the
different $f_0(1710)$ radiative decays under consideration is very
relevant, specially for the $\rho^0\gamma$ decay channel. The reason
why for the  $\rho^0\gamma$ channel the type-b loops is much larger than for
the $\omega\gamma$ and $\phi\gamma$ is that for the $\rho^0\gamma$ channel
the interference of the loop containing charged kaons with those
containing neutral kaons is constructive
(see table~\ref{tab:coeff_type_b}) for the mechanisms with a $\rho$
meson attached to the photon (the dominant one).
 On the contrary, for the
$\omega\gamma$ and $\phi\gamma$ decay channels this interference is
destructive.

In table~\ref{tab:results_others} we compare our result with other
theoretical determinations using quark models \cite{Close:2002sf,DeWitt:2003rs}.
\begin{table*}
\begin{center}
\begin{tabular}{|c|ccc|ccc||c|}
\hline
 &\multicolumn{3}{c|}{\cite{Close:2002sf}} & \multicolumn{3}{c|}{\cite{DeWitt:2003rs}} & present work \\ \hline\hline
& L & M & H &L & M & H &\\\hline
$f_0(1710)\to\rho^0\gamma$ &  42  &  94 & 705 &  24 &  $55^{+16}_{-14}$ & $410^{+200}_{-160}$ & $100\pm 40$ \\\hline
$f_0(1710)\to\phi\gamma$   & 800  & 718 &  78 & 450 &  $400^{+20}_{-20}$& $36^{+17}_{-14}$    & $15\pm 5$\\\hline
\end{tabular}
\end{center}
\caption{Comparison of the radiative decay widths into $\rho\gamma$ and $\phi\gamma$
with other theoretical predictions. All the units are in KeV.}
\label{tab:results_others}
\end{table*}
The L, M and H labels in the columns for the results of
refs.~\cite{Close:2002sf} and \cite{DeWitt:2003rs} refer to three
different versions of their model (light-weight, medium-weight and
heavy-weight glueball respectively).

We can see the large dispersion in the values obtained between the
different models and even within the same model in
refs.~\cite{Close:2002sf,DeWitt:2003rs}. Our results are of the same order
of magnitude than most of the results of the quark models. This means that
the meson loops play very important role in this decays and cannot be
neglected in realistic calculations.

Since we see that at least in the case of decay to $\rho \gamma$
the role of the vector mesons in the loop is relevant, this raises
the question of possible contributions of higher mass channels.
This is an issue that will have to be faced in the future. The
results of \cite{Albaladejo:2008qa}, where unfortunately no
information is provided on the coupling of the $f_0(1710)$ to the
different channels, should be relevant in this respect and the new
channels considered there in the scattering problem might provide
also a sizable contribution to the radiative decay. Hence, a
complete evaluation of the role of meson loops in the decay of the
$f_0(1710)$ resonance is not possible at the present time, but the
calculations carried out here, for the likely most important
channel, show clearly that the role of loops in the decay of this
resonance is something to deal with whenever one wishes to make
accurate evaluations of the radiative decays of that resonance to
draw conclusions on its nature. \\

\section{SUMMARY}

In summary, we have studied the relevance of the kaon loops  in the
radiative decay of the $f_0(1710)$ scalar resonance. Given the dominance of the
$K\bar K$ channel in the decay of this resonance, we have evaluated the
loops stemming from this decay which contain
either three pseudoscalar mesons (type-a in
fig.~\ref{fig:diag}) or two pseudoscalars and one vector meson
(type-b), in analogy to what was used in previous works regarding the
radiative decay of the $f_0(980)$ and $a_0(980)$ resonances.

By using arguments of gauge invariance, it can be shown that only type-a
and -b loops need to be evaluated and that type-a is convergent while the
logarithmic divergence of the type-b loop can be written in terms of the
well known two pseudoscalar mesons loop function.

The contribution of these meson loops to the radiative decay width is
very relevant and should be taken into account in the theoretical
calculations. With no intention to make a precise evaluation of the
role of meson loops, lacking information from the coupling of the
$f_0(1710)$  resonance to higher mass meson-meson channels, the
present work has the value of showing the order of magnitude of what
should be expected, which is sufficient to claim that their
consideration is important for a proper understanding of the
radiative decay widths. Experimental measurements of these radiative
decays would be mostly welcome to deepen into the understanding of
this controversial scalar resonance and could be accessible at
experimental facilities like BESIII.

\section*{Acknowledgments}
We thank financial support from MEC (Spain) grants No. FPA2004-03470,
FIS2006-03438, FPA2007-62777, Fundaci\'on S\'eneca grant No. 02975/PI/05,
and the Japan(JSPS)-Spain collaboration agreement.
One of the author (H.N.) is
the Research Fellow of the Japan Society for the Promotion of Science
(JSPS) and supported by the Grant for Scientific Research from JSPS
(No.~18$\cdot$8661).
This research is  part of the EU Integrated
Infrastructure Initiative Hadron Physics Project under contract
number RII3-CT-2004-506078.

\end{document}